\documentclass[useAMS,usenatbib]{mn2e}
\usepackage{graphicx}
\def\mch{M$\rm^{c}$Hardy\,}
\def\etal{et al.~\rm}
\def\xmm{{\em XMM-Newton}}
\def\ch{{\em Chandra }}

\begin{document}

\title[A 610~MHz Survey of the $1^H${\em XMM-Newton}$\slash${\em Chandra} Survey Field]
{A DEEP GMRT 610~MHz SURVEY OF THE $1^H$ {\em XMM-Newton}$\slash${\em Chandra} SURVEY FIELD}

\author[D Moss \etal]
{D. Moss,$^{1}$ \thanks{E-mail:dm@astro.soton.ac.uk}N. Seymour,$^{2}$ I. M. M$\rm^{c}$Hardy,$^{1}$ 
T. Dwelly,$^{1}$ M. J. Page,$^{3}$ N. S. Loaring $^{4}$ \\
$^{1}$ School of Physics and Astronomy, University of Southampton, Highfield, Southampton SO17 1BJ, UK\\
$^{2}$ Spitzer Science Centre, California
Institute of Technology, Pasadena, CA 91125, USA\\
$^{3}$ Mullard Space Science Laboratory, University College London,
Holmbury St Mary, Dorking, RH5 6NT, UK\\
$^{4}$ SAAO, PO Box 9, Observatory, 7935, South Africa\\
}

\date{\today}
\pagerange{\pageref{firstpage}--\pageref{lastpage}}
\pubyear{2006}

\label{firstpage}

\maketitle

\begin{abstract}
We present the results of a deep 610~MHz survey 
of 
the 
$1^H$ \xmm$\slash$\ch  
survey area with the GMRT. 
The resulting maps
have a resolution of $\sim$7 arcsec and an rms noise limit of 60~$\mu$Jy. 
To a 5$\sigma$ detection limit of 300~$\mu$Jy we detect 223
sources within a survey area of diameter 64 arcmin.  We compute the 610
MHz source counts and compare them to those measured at other radio
wavelengths. The well know flattening of the Euclidean-normalised 1.4~GHz 
source counts below $\sim$2 mJy, usually explained by a population
of starburst galaxies undergoing luminosity evolution, is seen 
at 610~MHz. The 610~MHz source counts can be modelled
by the same populations that explain the 1.4~GHz source
counts, assuming a spectral index of -0.7 for the starburst galaxies
and the steep spectrum AGN population.  We find a similar 
dependence of luminosity evolution on redshift for the starburst galaxies at 610MHz
as is found at 1.4~GHz (i.e. `$Q$'= $2.45^{+0.3}_{-0.4}$).
\end{abstract}


\section{Introduction}
\label{section:intro}
Deep radio surveys provide the basic observational data, i.e. counts of
the number of sources as a function of radio flux density, on which
our understanding of the evolution of the radio source population in
the universe is based (e.g. Longair 1966; Rowan-Robinson \etal
1993; Hopkins \etal 1998; Richards 2000; Seymour \etal 2004, Huynh \etal 2005). They have thus been the subject of considerable observational
effort, particularly at frequencies of 1.4~GHz and above where
angular resolution of a few arcsec, or better, is now commonplace and
source confusion is unimportant. Radio surveys are also forming an
increasingly important part of multi-waveband observational programmes
designed to study, in a more holistic manner, the broad-band evolution
of various classes of sources such as AGN and starburst galaxies. 
At low frequencies (i.e. below 1.4~GHz), where steep spectrum
sources such as starburst galaxies 
are expected to make a larger contribution relative to flatter spectrum AGN, 
radio surveys have, until recently, been
limited to relatively bright flux limits ($\sim 2.5$~mJy, Valentijn
1980).
This has been primarily due to the lack of a radio telescope with the
sensitivity and angular resolution required to reach to faint flux densities
without being limited by source confusion. However, the Giant Metre-wave Radio
Telescope (GMRT) of India's National Centre for Radio Astrophysics now provides
the necessary capabilities at low frequency.

In recent years, a number of deep radio surveys at 1.4~GHz have reached
$\mu$Jy levels (e.g. Hopkins \etal 1998; Richards 2000; Ivison \etal 2002; Seymour \etal 2004, Huynh \etal 2005).
They confirm the previous result (Windhorst \etal 1985) of an upturn of the 
Euclidean normalised differential source counts below
$\sim2$~mJy.  It is postulated that this upturn is due to a
population of sources other than AGN, 
and is commonly 
ascribed to
starburst galaxies (e.g. Condon 1989; 
Benn \etal 1993).
The source counts
have been modelled with a combination of an AGN population (e.g. Dunlop \&
Peacock 1990) and a starburst population, both undergoing luminosity
evolution (e.g. Hopkins \etal 1998). 

The previous deepest radio surveys at 610~MHz were made with the
Westerbork Synthesis Radio Telescope (WSRT) in the 1970s.
Katgert (1979) combined several such surveys
to compute source counts down to 22~mJy. Later WSRT surveys of galaxy
clusters achieved lower detection limits (e.g. 2.5~mJy, Valentijn
1980). However, none of these surveys was sufficiently deep to reveal
the upturn in the source counts.

Here, using GMRT observations, we present 
a deep 610~MHz survey 
of a 
survey field (at $1^h $, $-4^{\circ}$), 
which is one of two very deep survey fields
which we have observed in the X-ray, optical and IR bands, as well as
at 1.4~GHz with the VLA (e.g. 
\mch \etal 2003; Loaring \etal 2005; Seymour \etal 2004
). 


In this paper we derive the
610~MHz source counts to a flux density limit ($300 \mu$Jy) 
which is almost an order of magnitude fainter than the predicted upturn in the source counts.
In this paper the observations and data reduction are described in Section \ref{section:obs}. 
The construction of the catalogue is discussed in Section \ref{section:analysis}
and we present the computation and modelling of the 610~MHz source counts
in Sections \ref{section:sourcecounts}. 
Conclusions are presented in
Section \ref{section:conclusion}.

Throughout, we assume a concordance cosmology with $H_{0}=70 $kms$^{-1}$Mpc$^{-1}$, $\Omega_{M}=0.3$, 
and $\Omega_{\Lambda}=0.7$ (Spergel \etal 2003).

\section{Observations and Data Reduction}
\label{section:obs}
In August of 2004, the $1^H$ field (RA=$1^h 45^m 27^s$ and $Dec=-4^{\circ} 34\arcmin 42\arcsec$)
was observed for approximately 4.5 hours at 610~MHz with the 
GMRT. Observations were carried out in dual band, spectral line mode, the
former to maximise bandwidth and the latter to minimize chromatic aberration.
Two sidebands, each of 128 spectral channels of 125 kHz,
were centred on 602 and 618 MHz to give a total of 32 MHz bandwidth,
with two independent circular polarisations recorded. Visibilities were
recorded every 17 seconds. Observations were made of the phase calibration
source 0054+035 every 45 minutes. The bandpass and absolute flux
calibration were provided by several observations of 3C48.

All reduction was performed with the National Radio Astronomy Observatory's AIPS software. The two
sidebands were calibrated, preliminarily imaged and self calibrated independently and 
then combined in the $uv$ plane before final imaging.
The visibilities from each baseline and channel were carefully inspected and edited
to remove interference, 
cross-talk between antennas, 
instrumental glitches, and the poor quality data recorded at the
beginning of each pointing.

\subsection{Calibration}
\label{section:calib}

Absolute flux and bandpass calibration was performed in the standard way with
3C48. Phase calibration was performed using 0054+035. Each side band
underwent 1 cycle of phase self-calibration. We experimented with further 
iterations, but found that they did 
not improve the image quality, or reduce the noise significantly.
The two sidebands were then combined in the $uv$ plane using the customised AIPS task 
UVFLP\footnote{Provided by Dave Green, http://www.mrao.cam.ac.uk/$\sim$dag/}.

\subsection{Imaging Strategy}
\label{section:imaging}

Imaging was performed
over a five by five grid of 1024$\times$1024 arcsec facets, with the phase of the data
shifted to the tangent point at the centre of each facet. 
This strategy was chosen to ensure that no point
on any image would be more than 10 arcmin from a tangential pointing
position, minimizing the smearing of sources far from the phase tracking centre (due to the
breaking of the coplanar array assumption).

The 256 channels were combined into one $uv$
plane during the gridding process, imaged, and deconvolution performed with
the AIPS task IMAGR. The final 25 image facets used for analysis were
total intensity maps, with pixel size of 1 $\times$ 1
arcsec.
Made
with a natural weighting, the central rms noise level achieved was 60~$\mu$Jy
with a restoring beam of 7.3 $\times$ 6.1 arcsec.
Areas of slightly higher rms noise surround the brightest 5
sources, the slightly extended structures of which are imperfectly deconvolved.

Special care was taken to remove the effects of bright sources located within
the side lobes of the primary beam. Small images were made of all surrounding
sources found at 1.4~GHz in the NRAO VLA Sky Survey (NVSS, Condon \etal 1998), including 
them in the cleaning process to ensure that their
side lobes were properly removed from the science images.

The maps of the individual facets were combined into a final mosaic
using the AIPS task FLATN, with no correction made for the 
degradation of the primary beam sensitivity at this stage.

\section{Catalogue Construction}
\label{section:analysis}

\subsection{Source Extraction}
\label{section:SAD}

\begin{figure*}
\includegraphics[width=13cm,angle=0]{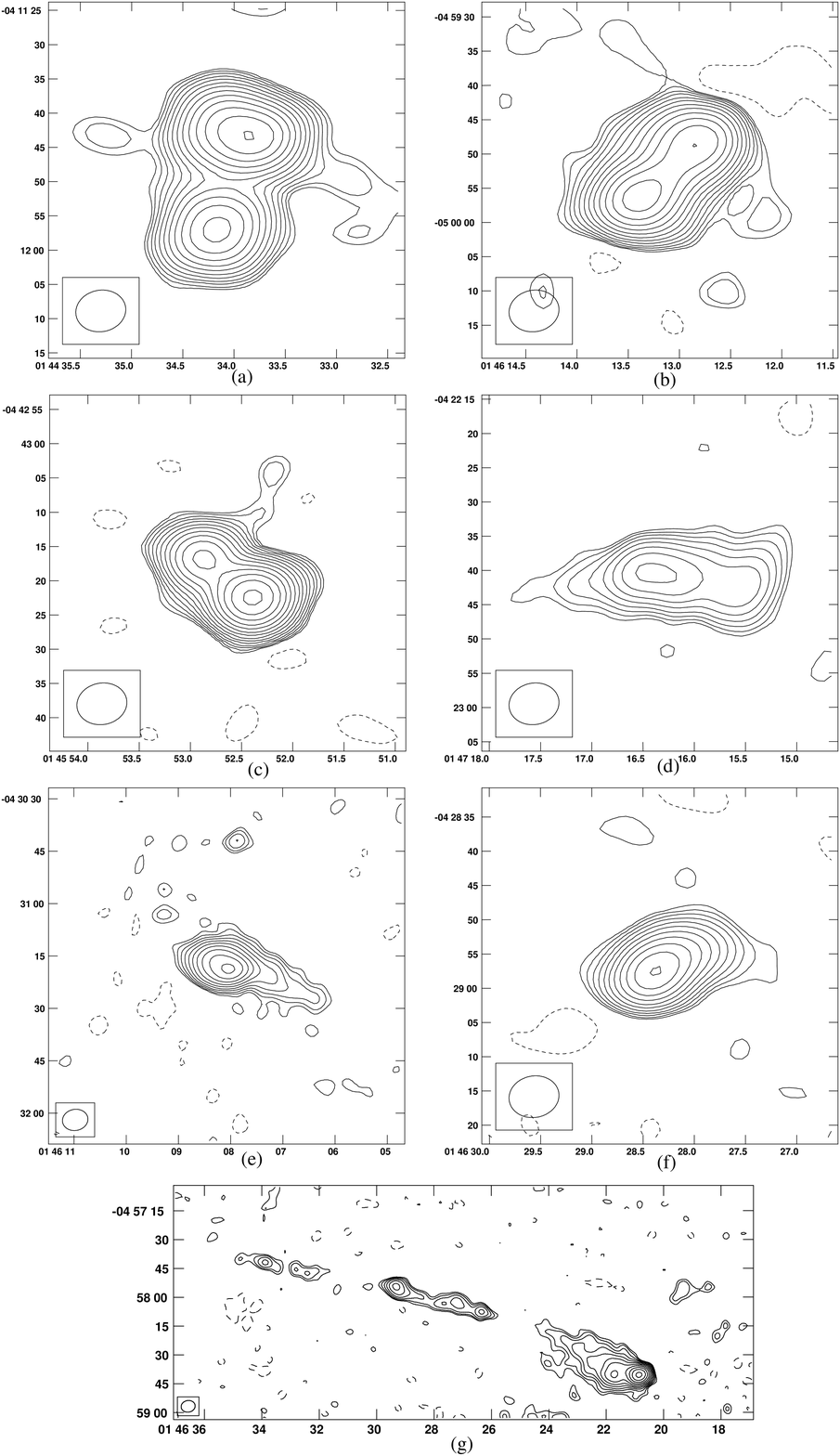}
\caption{
Plots of the contours of radio emission from sources 1, 5, 8, 16, 23 and 24 are shown
in panels (a), (b), (c), (d), (e), and (f) respectively. Panel (g) shows 
sources 10 (consisting of the two elongated patches of emission in the middle and top left)
and 20 (to the bottom right).
Contours are shown at -2, and at $2^{1+n/2}$ (where n=1,2,3..) times the local noise.}
\label{fig:contsoftvstat}
\end{figure*}

Sources were extracted 
with the
AIPS task SAD. 
A conservative peak flux density detection limit of $5\sigma$ (i.e. $300~\mu$Jy)
was used to minimize the number of noise spikes spuriously detected as
sources.  In the areas surrounding the five brightest sources, detection was performed 
separately with higher detection thresholds to account for the higher
rms noise.
Within the 20 percent power radius of the GMRT
primary beam at 610~MHz (32 arcmin), 213 sources were discovered above
a 5$\sigma$ peak flux density detection limit of 300~$\mu$Jy.  
In order to determine the success 
of the SAD source extraction, both the science images and the residual
noise maps were  carefully inspected.  
There were 8 extended sources where the Gaussian model fit
by SAD inadequately described the data. The characteristics of these
sources were determined using the AIPS task TVSTAT, and contour plots of them
are shown in Fig \ref{fig:contsoftvstat}.
Five of these appear to contain two peaks joined by extended emission,
i.e. double lobe sources.

Five further pairs of sources 
are separated by less than 20~arcsec, and could potentially be components of a double source.
However, 
inspection of $i'$ band imaging from SuprimeCam on the Subaru telescope, and
the Digitised Sky Survey (DSS) images 
shows that
for four of these cases, separate optical counterpart candidates are present
for the each source of radio emission, suggesting that they are unrelated.

The emission from the two sources constituting the remaining potential double 
is separated by 17~arcsec and shows no sign of extention in the direction 
linking the two. They have been included in the catalogue separately as sources
106 and 149.


To search for further low surface brightness extended sources, a low resolution map was made using 
a similar strategy to that presented in Section \ref{section:imaging}, but 
with the $uv$ visibility data tapered.
With a restoring beam of 12$\times$12 arcsec,
and an rms noise of 72~$\mu$Jy, ten further sources were discovered above
5$\sigma$, and added to the catalogue.

\subsection{Resolution}
\label{section:resolution}
\begin{figure}
\includegraphics[width=6cm,angle=0]{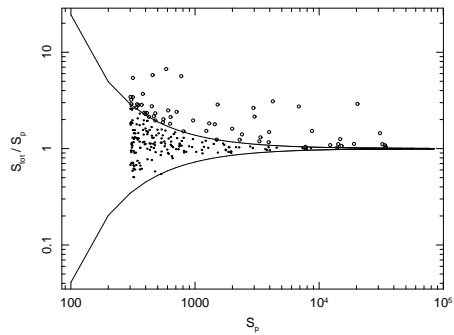}
\caption{
The ratio of $S_{tot} $to$ S_P$ as a function of $S_P$. Solid lines indicate the 
upper and lower envelopes of the flux ratio distribution used to define an
unresolved source. Open circles are those sources considered resolved, while dots
indicate those considered unresolved.
}
\label{fig:res_env}
\end{figure}

In the analysis of low signal to noise detections, noise spikes can 
cause Gaussian fitting routines such as SAD to poorly fit the width of a source, and hence inaccurately
measure the total flux density. The ratio of the total to peak flux density 
($S_{tot}/S_P$) is often used to probe this effect. In Fig. \ref{fig:res_env}, we plot 
$S_{tot}/S_P$ against $S_P$. Under the assumption that the measurement of any
source as having $S_{tot}$ $<$ $S_P$ is due to the effect of noise, 
we use the the distribution of $S_{tot}/S_P$ to define a criterion for 
a source to be considered resolved. We define an envelope that contains 98 percent of the sources
with $S_{tot}/S_P < 1$ as $S_{tot}/S_P = 0.88^{2.5/S_P}$, where $S_P$ is measured
in mJy.
This envelope is then mirrored about $S_{tot}/S_P = 1 $. Any source
within this envelope is considered unresolved, and its peak flux density taken
as the best measure of its true flux density.

\subsection{Flux Density Corrections}
\label{section:corrections}

There are a number of potential systematic effects which can cause erroneous 
measurements of flux density in deep radio observations. We evaluate the impact of each as follows.

\subsubsection{Bandwidth Smearing}
\label{section:band}
Bandwidth smearing is the result of channels of finite bandwidth
being used in an interferometric inversion that assumes monochromatic radiation.
Bandwidth smearing manifests itself as a
radial blurring
where the total flux density of a source is conserved, but the peak
flux density is reduced. The effect increases with distance from the phase tracking
centre. Our strategy of keeping the 128 channels
of each sideband separate until the $uv$ gridding process ensured 
minimal bandwidth smearing, as the 
bandwidth over which this effect is relevant is the channel width
of 125kHz rather than the total sideband bandwidth of 16MHz.
Following Chapter 18 of 
Taylor \etal (1999),
and approximating the $uv$
coverage to a Gaussian tapered circle, and the bandpass as square,
the maximal theoretical effect of bandwidth
smearing (at the edge
of the survey area) for our data is $\left| (S_{p}-S_{p}^{true})/S_{p}^{true} \right| < 0.0003$ where $S_{p}$ is
the measured peak intensity and $S_{p}^{true}$ is the true, unsmeared peak
flux density.
Thus, bandwidth smearing is not considered further.

\subsubsection{Time delay Smearing}
\label{section:time}
Sources rotate in the sky during the 17 second integration of each
of our visibilities. In the simplified case of a target at a celestial
pole, the effect is essentially a smearing of each source in a
direction tangential to that towards the phase tracking centre. In
this simplified case, and approximating to a circular $uv$ coverage
with Gaussian tapering (again following Taylor \etal 1999), the {\em maximum} 
theoretical effect at the edge
of our survey area is a decrease of peak flux density of 4\%, which for the majority of sources
is smaller than the flux density measurement uncertainties.

\begin{figure}
\includegraphics[width=6cm,angle=0]{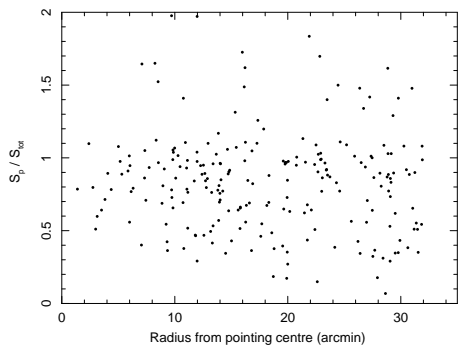}
\caption{The variation of $S_{p}/S_{tot}$ across the primary beam.
Any time or bandwidth smearing effecting the data would manifest itself as a decline of the 
$S_{p}$ to $S_{tot}$ ratio with radius, which is not seen.}
\label{fig:bandtsmear}
\end{figure}

Fig. \ref{fig:bandtsmear} shows the ratio of measured peak to measured 
total flux density of the observed sources.
There is no apparent dependence of this ratio on radial distance from the phase tracking centre, 
as would be apparent should either of these smearing effects be important.
A Pearson's correlation test gives a value of -0.07, implying 
no significant
correlation.
This result confirms that neither bandwidth
nor time delay smearing has a significant effect on the quality
of our images, and hence neither effect was considered further.

\subsubsection{3D Smearing}
\label{section:3d}
\begin{figure}
\includegraphics[width=6cm,angle=0]{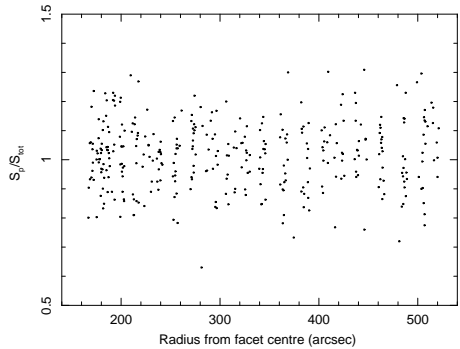}
\caption{The ratio of $S_p$ to $S_{tot}$ for simulated sources, plotted against radial distance from the
local facet tangent pointing centre at radii where any effect, if present, would be most noticeable. 
If 3D smearing were important, 
$S_{p}/S_{tot}$ would show a decrease with radius from the local pointing centre.}
\label{fig:facet}
\end{figure}

Sources far from the tangential
pointing have their flux smeared out due to the inaccuracy of the
assumption of 
a coplanar array
in the interferometric Fourier transform
relation. Our imaging strategy was expected to minimize this effect, but
simulations were carried out to discover the magnitude of any remaining 3D smearing.
A series of fake point sources were inserted into our $uv$
coverage at a constant radius from the phase tracking centre of the field. This strategy
isolates 3D smearing across a facet from the radially dependent bandwidth and time
delay smearing across the primary beam. This modified $uv$ data was then imaged with the 
same strategy as used for the
science images. The ratio of $S_p$ to $S_{tot}$ of these fake sources showed no
dependence on distance from the local facet tangential pointing centre,
as seen in Fig. \ref{fig:facet}. A Pearson's correlation test gives 0.03, indicative of 
no significant correlation.  
This result confirms that 3D smearing is not important, and so it is not considered further.

\subsubsection{Clean Bias}
\label{section:clean}
\begin{figure}
\includegraphics[width=6cm,angle=0]{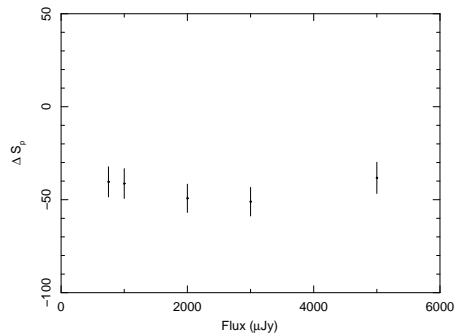}
\caption{The average measured decrease in total flux density, 
for model sources inserted into the $uv$ data, imaged, and detected by SAD, 
as a function of peak flux density. Error bars give the uncertainty on the 
determination of the mean.
}
\label{fig:cleanbias}
\end{figure}

The Clean algorithm used to deconvolve the dirty beam can, in presence of noise,
reduce the measured flux of sources. The superposition of a noise spike with 
a sidelobe from a real source can lead to the placing of clean components 
at positions not occupied by real sources. The removal of the dirty beam from this
position can then remove flux from real sources, and lead to systematic underestimation
of their flux densities (Condon \etal 1998; Prandoni \etal 2000a; Huynh \etal 2005). 
To quantify the effect of this 
Clean bias on our measurements, a series of fake sources with known flux densities
were inserted randomly into the full $uv$ data at positions not 
coincident with any real sources, and away from the areas of high noise surrounding the
brightest sources. Imaging was then conducted in a manner similar to 
that used for the science images, and source extraction repeated. 
The measured flux densities of these model sources were then compared to the input model 
source parameters and the average offset as a function of 
flux is shown in Fig. \ref{fig:cleanbias}. 
The average measured offsets are consistent within the errors on the determination of the mean
at each flux.
The fitting of a simple constant to the offsets 
at these fluxes gives 44~$\mu$Jy as the amount by which the imaging, cleaning and source
extraction processes have reduced the true model source flux densities. 
The flux densities of the real sources have been corrected thus, before any correction
for the degradation in sensitivity of the primary beam.

\subsubsection{Primary Beam Attenuation}
\label{section:pb}
Any interferometric observation is
modulated by the primary beam pattern of the array elements. The
flux densities of all objects were corrected for the GMRT 610~MHz primary
beam response using an eighth order polynomial provided by N.
Kantharia and A. Pramesh Rao (Addendum to GMRT Technical Report
R00185).

\subsection{Catalogue}
\label{section:cat}

The full
catalogue is given in Table \ref{tab:catalogue}, in
order of descending flux density, giving: 
(1) source number; 
(2) right ascension (J2000);
(3) declination (J2000); 
(4) detection signal to noise ratio; 
(5) peak flux density, $S_{p}$ in $\mu$Jy beam$^{-1}$;
(6) angular size in arcsec (as determined from the deconvolution 
of the clean beam by SAD, with the exception of the 8 extended sources whose 
flux densities were extracted using TVSTAT, in which case we give the 
distance in arcsec between the 4$\sigma$ contours along the line of greatest extent) for those 
sources considered reliably resolved; 
(7) integrated flux density, $S_{tot}$ in $\mu$Jy for those sources considered reliably resolved; 
(8) 1$\sigma$ error on $S_{p}$ for point sources, or $S_{tot}$ for those resolved.


\begin{table*}
\caption{The complete catalogue of sources. Source Number, RA, Dec,
detection signal to noise ratio (SNR), 
peak flux density $S_p$ in $\mu$Jy beam$^{-1}$,
angular size in arcsec (see Section \ref{section:cat}), 
integrated flux density $S_{tot}$ in $\mu$Jy, both for sources considered reliably resolved,
and $\sigma_{S}$, the error on the best determination of the flux density, in $\mu$Jy.}
\label{tab:catalogue}
\begin{tabular}{lccccccc}
Source & Right Ascension & Declination & SNR & peak  & $\theta$ & $S_{tot}$ & $\sigma$   \\ 
 &  &  &  & ($\mu$Jy beam$^{-1}$) & (\arcsec) & ($\mu$Jy)  & ($\mu$Jy ) \\ 
1   $^{\dagger}$  &    01:44:34.0 &  -04:11:50.2  &    338.9    &    55383.1 &   32.0   &     160804.3  &     450.6 \\ 
2     &    01:47:15.9 &  -04:50:09.8  &    537.8    &    138106.8 &   3.73   &     153572.6  &     285.5 \\ 
3     &    01:44:44.3 &  -04:55:34.9  &    562.6    &    72536.4 &   3.67   &     78817.5  &     140.0 \\ 
4     &    01:45:40.5 &  -04:12:35.0  &    570.5    &    68278.8 &   2.29   &     71533.8  &     125.3 \\ 
5   $^{\dagger}$  &    01:46:13.1 &  -04:59:53.4  &    114.2    &    20786.1 &   19.51   &     56590.3  &     453.9 \\ 
6     &    01:43:49.2 &  -04:18:17.9  &    237.2    &    50278.1 &   3.29   &     56077.4  &     236.3 \\ 
7     &    01:47:21.1 &  -04:45:41.6  &    204.4    &    48648.2 &   3.07   &     52916.9  &     258.8 \\ 
8   $^{\dagger}$  &    01:45:52.6 &  -04:43:19.1  &    515.1    &    35946.5 &   5.16   &     51873.4  &     99.0 \\ 
9     &    01:45:57.8 &  -05:00:26.4  &    252.4    &    42551.8 &   2.23   &     45389.3  &     179.7 \\ 
10   $^{\dagger}$  &    01:46:21.3 &  -04:58:39.7  &    70.5    &    12659.7 &   21.23   &     38896.4  &     505.1 \\ 
11     &    01:45:20.3 &  -05:02:00.9  &    126.9    &    22403.5 &                &     &                          176.4 \\ 
12     &    01:45:32.8 &  -04:30:54.0  &    318.3    &    19557.7 &   3.07   &     21884.5  &     68.7 \\ 
13     &    01:45:37.7 &  -04:34:00.8  &    246.2    &    14965.6 &   4.27   &     18768.6  &     74.9 \\ 
14     &    01:46:09.6 &  -04:55:3.0  &    143.1    &    17914.9 &                &     &                          125.1 \\ 
15     &    01:44:44.8 &  -04:40:26.8  &    232.8    &    16903.4 &   2.5   &     17331.7  &     74.4 \\ 
16   $^{\dagger}$  &    01:47:16.3 &  -04:22:40.7  &    25.3    &    5728.9 &   12.71   &     16109.8  &     593.5 \\ 
17     &    01:44:43.0 &  -04:07:45.6  &    65.3    &    13582.1 &   3.13   &     15794.5  &     241.6 \\ 
18     &    01:45:43.7 &  -04:43:36.3  &    146.0    &    9982.9 &   7.01   &     15183.6  &     100.5 \\ 
19     &    01:44:14.7 &  -04:45:06.5  &    132.4    &    14394.5 &                &     &                          108.7 \\ 
20   $^{\dagger}$  &    01:46:29.2 &  -04:57:55.4  &    12.8    &    2516.3 &   10.24   &     13578.0  &     1000.8 \\ 
21     &    01:46:15.4 &  -04:40:29.6  &    129.3    &    9875.9 &   1.74   &     10289.8  &     79.5 \\ 
22     &    01:45:01.1 &  -04:59:47.7  &    61.3    &    9650.2 &                &     &                          157.2 \\ 
23   $^{\dagger}$  &    01:46:08.1 &  -04:31:17.9  &    49.2    &    3490.1 &   10.51   &     9157.1  &     173.5 \\ 
24   $^{\dagger}$  &    01:46:28.3 &  -04:28:57.1  &    65.8    &    5696.1 &   7.38   &     8420.8  &     123.6 \\ 
25     &    01:46:06.1 &  -04:35:44.2  &    122.2    &    8362.8 &                &     &                          68.4 \\ 
26     &    01:47:06.7 &  -04:53:31.5  &    16.0    &    4292.4 &   10.6   &     8233.8  &     487.9 \\ 
27     &    01:47:28.4 &  -04:41:37.2  &    20.5    &    5334.7 &   7.04   &     8058.5  &     379.6 \\ 
28     &    01:44:05.8 &  -04:24:36.6  &    9.7    &    1273.8 &   27.92   &     8019.3  &     743.1 \\ 
29     &    01:45:14.4 &  -04:12:00.1  &    62.9    &    7892.5 &                &     &                          125.4 \\ 
30     &    01:45:20.3 &  -04:22:58.3  &    50.2    &    3671.0 &   8.67   &     7823.5  &     147.9 \\ 
31     &    01:46:36.9 &  -04:43:39.7  &    75.7    &    7707.5 &                &     &                          101.8 \\ 
32     &    01:47:11.0 &  -04:50:18.3  &    30.9    &    7343.6 &                &     &                          237.1 \\ 
33     &    01:47:25.3 &  -04:43:23.2  &    28.0    &    6987.8 &                &     &                          249.2 \\ 
34     &    01:47:04.7 &  -04:49:29.2  &    32.0    &    6386.3 &                &     &                          199.2 \\ 
35     &    01:45:02.3 &  -04:12:38.6  &    48.6    &    6115.7 &                &     &                          125.8 \\ 
36     &    01:46:11.8 &  -04:16:21.7  &    33.1    &    3825.1 &   8.33   &     6099.6  &     177.6 \\ 
37     &    01:43:54.2 &  -04:18:25.2  &    24.9    &    4888.8 &   3.78   &     6034.9  &     238.0 \\ 
38     &    01:43:54.4 &  -04:47:44.4  &    35.6    &    5929.3 &                &     &                          166.3 \\ 
39     &    01:45:10.0 &  -04:24:22.8  &    56.0    &    4010.5 &   4.52   &     5236.4  &     91.8 \\ 
40     &    01:45:52.6 &  -05:05:24.3  &    10.4    &    2914.3 &   7.85   &     5122.7  &     468.1 \\ 
41     &    01:46:41.0 &  -04:25:12.2  &    46.3    &    5079.1 &                &     &                          109.6 \\ 
42     &    01:45:04.4 &  -04:03:18.8  &    17.0    &    4885.3 &                &     &                          287.0 \\ 
43     &    01:46:12.3 &  -04:37:48.5  &    54.8    &    3990.4 &   3.8   &     4732.0  &     86.2 \\ 
44     &    01:45:45.1 &  -04:58:54.7  &    32.5    &    4681.4 &                &     &                          143.6 \\ 
45     &    01:46:18.4 &  -04:19:28.8  &    7.5    &    854.7 &   19.57   &     4588.2  &     548.6 \\ 
46     &    01:44:49.1 &  -04:31:43.9  &    66.5    &    4584.5 &                &     &                          68.8 \\ 
47     &    01:46:22.2 &  -04:10:09.1  &    23.5    &    4581.7 &                &     &                          194.5 \\ 
48     &    01:44:31.4 &  -04:36:49.5  &    39.7    &    3148.9 &   6.58   &     4405.4  &     108.8 \\ 
49     &    01:45:46.3 &  -04:13:41.0  &    37.7    &    4366.7 &                &     &                          115.6 \\ 
50     &    01:45:19.8 &  -04:26:20.3  &    65.0    &    4339.5 &                &     &                          66.6 \\ 
51     &    01:47:04.1 &  -04:50:55.8  &    18.9    &    4062.0 &                &     &                          214.9 \\ 
52     &    01:44:10.8 &  -04:48:12.2  &    31.1    &    4057.3 &                &     &                          130.3 \\ 
53     &    01:44:31.9 &  -04:43:27.4  &    24.0    &    2112.3 &   7.94   &     3731.1  &     150.1 \\ 
54     &    01:44:32.5 &  -04:37:44.6  &    38.1    &    3010.3 &   3.61   &     3705.8  &     95.5 \\ 
55     &    01:47:19.8 &  -04:42:1.0  &    5.0    &    1180.7 &   12.71   &     3694.7  &     677.8 \\ 
56     &    01:45:09.3 &  -04:17:33.5  &    21.6    &    2042.5 &   7.17   &     3679.3  &     161.3 \\ 
57     &    01:44:46.3 &  -05:02:24.7  &    5.6    &    1376.2 &   11.98   &     3677.6  &     601.7 \\ 
58     &    01:46:15.2 &  -05:00:27.2  &    5.0    &    1128.4 &   14.81   &     3312.9  &     602.0 \\ 
59     &    01:46:36.5 &  -04:24:58.0  &    31.4    &    3300.8 &                &     &                          104.9 \\ 
60     &    01:47:33.0 &  -04:32:01.3  &    6.1    &    1127.7 &   23.87   &     2991.8  &     540.3 \\ 
61     &    01:45:0.0 &  -04:11:20.3  &    20.2    &    2878.9 &                &     &                          141.9 \\ 
62     &    01:46:19.5 &  -04:08:40.9  &    13.0    &    2841.9 &                &     &                          217.3 \\ 
\multicolumn{8}{l}{$^{\dagger}$ Source parameters extracted using TVSTAT}
\end{tabular}
\end{table*}

\begin{table*}
\contcaption{}
\begin{tabular}{lccccccc}
Source & Right Ascension & Declination & SNR & peak  & $\theta$ & $S_{tot}$ & $\sigma$   \\ 
 &  &  &  & ($\mu$Jy beam$^{-1}$) & (\arcsec) & ($\mu$Jy)  & ($\mu$Jy ) \\ 
63     &    01:46:40.3 &  -04:30:40.0  &    5.2    &    575.8 &   20.76   &     2804.8  &     485.0 \\ 
64     &    01:45:39.5 &  -05:00:43.3  &    6.8    &    1230.9 &   19.37   &     2728.2  &     453.5 \\ 
65     &    01:44:43.3 &  -04:07:55.7  &    12.5    &    2674.2 &                &     &                          214.1 \\ 
66     &    01:45:27.6 &  -04:37:41.3  &    7.9    &    1418.4 &   14.76   &     2663.0  &     384.4 \\ 
67     &    01:46:02.8 &  -04:37:18.6  &    6.4    &    1176.6 &   18.44   &     2614.1  &     456.9 \\ 
68     &    01:46:53.8 &  -04:36:42.2  &    6.6    &    1195.6 &   17.27   &     2589.6  &     444.6 \\ 
69     &    01:44:29.9 &  -04:20:38.9  &    6.3    &    728.7 &   13.99   &     2490.6  &     362.6 \\ 
70     &    01:44:50.1 &  -04:46:22.7  &    28.6    &    2357.4 &                &     &                          82.3 \\ 
71     &    01:46:35.5 &  -04:32:13.6  &    25.4    &    2339.7 &                &     &                          91.9 \\ 
72     &    01:44:33.1 &  -04:05:45.3  &    7.7    &    2334.2 &                &     &                          302.1 \\ 
73     &    01:46:21.8 &  -04:34:42.5  &    29.3    &    2310.7 &                &     &                          78.6 \\ 
74     &    01:45:59.7 &  -04:28:57.2  &    32.6    &    2277.9 &                &     &                          69.8 \\ 
75     &    01:46:20.8 &  -04:24:26.5  &    24.8    &    2245.6 &                &     &                          90.3 \\ 
76     &    01:46:15.5 &  -04:40:15.4  &    11.8    &    950.8 &   10.81   &     2210.5  &     175.8 \\ 
77     &    01:45:22.1 &  -04:22:47.6  &    29.5    &    2190.5 &                &     &                          74.1 \\ 
78     &    01:46:29.5 &  -04:46:50.2  &    19.6    &    2078.0 &                &     &                          105.6 \\ 
79     &    01:47:10.9 &  -04:42:28.5  &    11.2    &    2037.8 &                &     &                          177.6 \\ 
80     &    01:43:31.5 &  -04:35:51.3  &    8.8    &    1913.4 &                &     &                          216.3 \\ 
81     &    01:47:20.4 &  -04:48:32.2  &    6.4    &    1886.5 &                &     &                          279.2 \\ 
82     &    01:46:31.8 &  -04:33:35.0  &    21.2    &    1869.6 &                &     &                          87.9 \\ 
83     &    01:46:34.2 &  -04:33:44.5  &    7.4    &    716.5 &   14.5   &     1859.3  &     232.2 \\ 
84     &    01:44:40.9 &  -04:51:01.1  &    5.8    &    673.8 &   19.35   &     1770.8  &     283.8 \\ 
85     &    01:45:34.2 &  -04:19:00.1  &    13.4    &    1185.0 &   5.86   &     1759.3  &     127.2 \\ 
86     &    01:45:4.0 &  -05:01:42.8  &    8.9    &    1747.4 &                &     &                          193.9 \\ 
87     &    01:47:00.8 &  -04:34:13.7  &    12.4    &    1695.7 &                &     &                          136.1 \\ 
88     &    01:47:22.3 &  -04:39:01.3  &    7.4    &    1681.9 &                &     &                          221.6 \\ 
89     &    01:46:0.0 &  -04:26:13.3  &    10.5    &    814.9 &   8.55   &     1675.6  &     150.0 \\ 
90     &    01:45:52.8 &  -04:31:50.6  &    10.1    &    695.0 &   9.59   &     1662.2  &     153.9 \\ 
91     &    01:47:21.7 &  -04:48:40.2  &    5.2    &    1627.1 &                &     &                          299.1 \\ 
92     &    01:43:29.3 &  -04:34:40.3  &    6.8    &    1603.8 &                &     &                          233.6 \\ 
93     &    01:43:31.1 &  -04:34:44.0  &    7.2    &    1598.4 &                &     &                          222.4 \\ 
94     &    01:45:39.3 &  -04:17:21.7  &    7.7    &    772.3 &   10.49   &     1596.5  &     194.6 \\ 
95     &    01:47:12.9 &  -04:18:27.8  &    5.5    &    1580.5 &                &     &                          283.6 \\ 
96     &    01:44:48.0 &  -04:23:13.9  &    7.9    &    703.1 &   8.8   &     1557.4  &     184.2 \\ 
97     &    01:46:40.0 &  -04:30:04.7  &    5.5    &    610.1 &   12.05   &     1503.8  &     251.4 \\ 
98     &    01:44:28.8 &  -05:00:52.6  &    5.9    &    1503.4 &                &     &                          236.8 \\ 
99     &    01:46:00.1 &  -04:44:57.9  &    9.2    &    753.8 &   8.3   &     1467.1  &     150.5 \\ 
100     &    01:46:37.3 &  -04:59:47.8  &    5.2    &    1429.2 &                &     &                          256.1 \\ 
101     &    01:46:05.7 &  -04:15:14.1  &    11.5    &    1404.0 &                &     &                          118.9 \\ 
102     &    01:46:16.3 &  -04:39:58.9  &    9.1    &    753.0 &   8.12   &     1363.5  &     141.9 \\ 
103     &    01:45:01.9 &  -04:44:55.3  &    5.2    &    433.0 &   11.73   &     1355.6  &     238.4 \\ 
104     &    01:47:17.1 &  -04:35:01.8  &    6.8    &    1344.8 &                &     &                          196.4 \\ 
105     &    01:46:04.7 &  -05:02:57.0  &    5.2    &    1327.7 &                &     &                          251.9 \\ 
106     &    01:44:28.9 &  -04:52:11.8  &    9.9    &    1305.4 &                &     &                          131.5 \\ 
107     &    01:45:49.6 &  -04:40:50.9  &    19.1    &    1301.9 &                &     &                          68.1 \\ 
108     &    01:45:31.0 &  -04:48:40.4  &    15.7    &    1282.7 &                &     &                          81.5 \\ 
109     &    01:43:58.7 &  -04:22:23.5  &    8.0    &    1281.2 &                &     &                          160.6 \\ 
110     &    01:47:15.3 &  -04:24:21.8  &    5.5    &    1281.1 &                &     &                          226.2 \\ 
111     &    01:45:53.2 &  -04:31:24.8  &    19.0    &    1271.1 &                &     &                          66.7 \\ 
112     &    01:45:58.8 &  -05:01:00.5  &    6.3    &    1258.2 &                &     &                          192.0 \\ 
113     &    01:45:24.7 &  -04:47:28.8  &    16.0    &    1244.8 &                &     &                          76.9 \\ 
114     &    01:45:52.6 &  -04:49:07.1  &    6.8    &    1241.8 &                &     &                          207.4 \\ 
115     &    01:45:38.3 &  -04:20:28.7  &    5.1    &    462.7 &   9.96   &     1232.1  &     224.3 \\ 
116     &    01:45:58.6 &  -04:14:52.6  &    10.1    &    1221.5 &                &     &                          120.3 \\ 
117     &    01:44:26.4 &  -04:39:23.8  &    6.7    &    1220.1 &                &     &                          210.8 \\ 
118     &    01:46:46.4 &  -04:13:43.7  &    5.3    &    1219.1 &                &     &                          229.7 \\ 
119     &    01:45:10.1 &  -04:15:22.3  &    11.0    &    1203.9 &                &     &                          108.9 \\ 
120     &    01:45:02.9 &  -05:03:35.2  &    6.6    &    1201.0 &                &     &                          211.2 \\ 
121     &    01:47:16.8 &  -04:44:22.5  &    5.0    &    1193.2 &                &     &                          223.8 \\ 
122     &    01:46:17.3 &  -04:33:52.9  &    7.0    &    577.5 &   8.76   &     1172.1  &     156.5 \\ 
123     &    01:45:35.8 &  -04:37:49.2  &    6.4    &    1165.7 &                &     &                          214.9 \\ 
124     &    01:45:23.7 &  -04:07:44.6  &    6.3    &    1160.3 &                &     &                          209.0 \\ 
125     &    01:45:49.0 &  -04:58:39.9  &    7.5    &    1155.7 &                &     &                          151.4 \\ 
126     &    01:44:39.5 &  -04:37:58.1  &    14.9    &    1142.9 &                &     &                          76.9 \\ 
\end{tabular}
\end{table*}

\begin{table*}
\contcaption{}
\begin{tabular}{lccccccc}
Source & Right Ascension & Declination & SNR & peak  & $\theta$ & $S_{tot}$ & $\sigma$   \\ 
 &  &  &  & ($\mu$Jy beam$^{-1}$) & (\arcsec) & ($\mu$Jy)  & ($\mu$Jy ) \\ 
127     &    01:46:52.3 &  -04:29:29.7  &    8.8    &    1114.2 &                &     &                          125.7 \\ 
128     &    01:46:19.0 &  -04:40:36.1  &    13.4    &    1113.5 &                &     &                          81.6 \\ 
129     &    01:46:59.1 &  -04:36:44.6  &    7.9    &    1084.0 &                &     &                          136.6 \\ 
130     &    01:45:21.2 &  -05:01:21.9  &    5.7    &    1069.4 &                &     &                          187.5 \\ 
131     &    01:46:08.3 &  -04:49:18.0  &    10.8    &    1064.2 &                &     &                          98.5 \\ 
132     &    01:46:58.8 &  -04:36:58.8  &    7.7    &    1052.1 &                &     &                          134.7 \\ 
133     &    01:45:25.1 &  -04:52:02.7  &    10.9    &    1049.1 &                &     &                          95.9 \\ 
134     &    01:46:45.2 &  -04:32:20.1  &    9.6    &    1045.8 &                &     &                          105.7 \\ 
135     &    01:45:12.7 &  -04:57:56.6  &    7.1    &    1022.4 &                &     &                          142.6 \\ 
136     &    01:45:16.9 &  -04:54:28.9  &    9.1    &    1017.6 &                &     &                          109.0 \\ 
137     &    01:44:51.8 &  -05:00:29.9  &    5.1    &    1015.0 &                &     &                          199.4 \\ 
138     &    01:44:09.8 &  -04:40:45.7  &    8.9    &    1011.2 &                &     &                          109.1 \\ 
139     &    01:45:15.3 &  -04:50:54.1  &    10.6    &    983.8 &                &     &                          91.6 \\ 
140     &    01:45:09.2 &  -04:47:46.8  &    11.9    &    977.4 &                &     &                          80.8 \\ 
141     &    01:44:46.3 &  -04:42:05.6  &    12.4    &    971.1 &                &     &                          78.2 \\ 
142     &    01:45:09.4 &  -04:43:31.0  &    13.5    &    970.9 &                &     &                          71.9 \\ 
143     &    01:46:04.2 &  -04:40:41.1  &    12.9    &    963.6 &                &     &                          74.5 \\ 
144     &    01:45:44.1 &  -04:58:46.1  &    6.0    &    937.6 &                &     &                          155.8 \\ 
145     &    01:47:08.3 &  -04:42:17.2  &    5.0    &    927.1 &                &     &                          185.4 \\ 
146     &    01:46:04.8 &  -04:21:27.7  &    10.0    &    922.6 &                &     &                          91.6 \\ 
147     &    01:45:16.3 &  -04:30:27.2  &    14.1    &    921.2 &                &     &                          65.3 \\ 
148     &    01:46:04.3 &  -04:39:05.6  &    12.2    &    894.3 &                &     &                          73.2 \\ 
149     &    01:44:29.9 &  -04:52:21.5  &    6.5    &    888.6 &                &     &                          135.9 \\ 
150     &    01:46:06.7 &  -04:24:49.9  &    10.6    &    882.3 &                &     &                          81.7 \\ 
151     &    01:44:54.2 &  -04:10:25.0  &    5.0    &    875.0 &                &     &                          166.5 \\ 
152     &    01:45:17.5 &  -04:34:15.9  &    13.5    &    866.8 &                &     &                          63.9 \\ 
153     &    01:46:13.5 &  -04:20:31.6  &    8.4    &    865.7 &                &     &                          99.9 \\ 
154     &    01:44:45.2 &  -04:33:39.6  &    10.8    &    799.2 &                &     &                          74.1 \\ 
155     &    01:45:49.1 &  -04:49:07.1  &    8.9    &    795.2 &                &     &                          89.3 \\ 
156     &    01:45:3.0 &  -04:34:15.2  &    11.7    &    783.5 &                &     &                          66.9 \\ 
157     &    01:47:01.2 &  -04:29:01.1  &    5.0    &    777.9 &                &     &                          145.3 \\ 
158     &    01:44:55.7 &  -04:12:17.3  &    5.1    &    775.7 &                &     &                          149.7 \\ 
159     &    01:44:51.1 &  -04:56:24.7  &    5.2    &    765.7 &                &     &                          147.6 \\ 
160     &    01:44:45.5 &  -04:27:49.5  &    9.5    &    754.6 &                &     &                          79.0 \\ 
161     &    01:45:26.9 &  -04:47:25.1  &    9.4    &    752.5 &                &     &                          79.6 \\ 
162     &    01:46:3.0 &  -04:24:09.4  &    8.7    &    736.1 &                &     &                          83.9 \\ 
163     &    01:45:45.2 &  -04:12:19.5  &    5.2    &    734.5 &                &     &                          141.2 \\ 
164     &    01:44:38.5 &  -04:45:10.6  &    7.9    &    729.9 &                &     &                          92.1 \\ 
165     &    01:43:58.4 &  -04:33:39.0  &    5.4    &    723.2 &                &     &                          129.2 \\ 
166     &    01:45:40.3 &  -04:49:5.0  &    8.2    &    721.0 &                &     &                          84.4 \\ 
167     &    01:45:40.5 &  -04:16:52.3  &    6.9    &    713.9 &                &     &                          100.5 \\ 
168     &    01:44:52.1 &  -04:55:20.6  &    5.2    &    711.7 &                &     &                          130.1 \\ 
169     &    01:45:04.4 &  -04:32:59.9  &    10.6    &    711.4 &                &     &                          67.2 \\ 
170     &    01:46:52.6 &  -04:41:48.6  &    5.1    &    704.2 &                &     &                          138.4 \\ 
171     &    01:46:43.7 &  -04:28:58.8  &    5.9    &    689.7 &                &     &                          115.3 \\ 
172     &    01:44:34.2 &  -04:26:47.4  &    7.4    &    672.7 &                &     &                          90.2 \\ 
173     &    01:45:03.8 &  -04:48:14.9  &    7.6    &    671.2 &                &     &                          87.6 \\ 
174     &    01:45:26.9 &  -04:36:05.2  &    10.2    &    658.5 &                &     &                          63.7 \\ 
175     &    01:45:35.1 &  -04:40:22.8  &    9.4    &    641.7 &                &     &                          67.9 \\ 
176     &    01:44:30.6 &  -04:34:53.1  &    7.3    &    631.5 &                &     &                          85.8 \\ 
177     &    01:45:44.2 &  -04:16:37.2  &    5.7    &    618.5 &                &     &                          102.3 \\ 
178     &    01:46:35.9 &  -04:24:21.5  &    5.1    &    609.2 &                &     &                          118.5 \\ 
179     &    01:44:48.5 &  -04:32:57.4  &    8.1    &    600.3 &                &     &                          72.6 \\ 
180     &    01:45:42.9 &  -04:36:29.8  &    8.9    &    595.7 &                &     &                          65.7 \\ 
181     &    01:46:45.4 &  -04:34:58.0  &    5.1    &    593.5 &                &     &                          107.7 \\ 
182     &    01:44:22.3 &  -04:32:55.7  &    6.1    &    588.1 &                &     &                          95.4 \\ 
183     &    01:46:03.1 &  -04:35:48.6  &    7.9    &    578.1 &                &     &                          73.1 \\ 
184     &    01:44:31.2 &  -04:37:02.2  &    6.4    &    563.5 &                &     &                          86.4 \\ 
185     &    01:45:56.0 &  -04:51:32.7  &    5.1    &    553.0 &                &     &                          107.6 \\ 
186     &    01:44:40.9 &  -04:40:03.2  &    6.6    &    545.2 &                &     &                          82.3 \\ 
187     &    01:45:31.5 &  -04:23:36.8  &    6.8    &    533.7 &                &     &                          74.4 \\ 
188     &    01:46:12.0 &  -04:22:27.5  &    5.3    &    527.2 &                &     &                          96.1 \\ 
189     &    01:44:38.8 &  -04:27:44.3  &    5.9    &    516.5 &                &     &                          83.4 \\ 
190     &    01:44:46.8 &  -04:47:22.0  &    5.3    &    512.4 &                &     &                          96.8 \\ 
\end{tabular}
\end{table*}

\begin{table*}
\contcaption{}
\begin{tabular}{lccccccc}
Source & Right Ascension & Declination & SNR & peak  & $\theta$ & $S_{tot}$ & $\sigma$   \\ 
 &  &  &  & ($\mu$Jy beam$^{-1}$) & (\arcsec) & ($\mu$Jy)  & ($\mu$Jy ) \\ 
191     &    01:44:46.4 &  -04:48:09.8  &    5.0    &    509.7 &                &     &                          100.3 \\ 
192     &    01:45:31.3 &  -04:29:29.9  &    7.2    &    499.9 &                &     &                          68.5 \\ 
193     &    01:46:02.6 &  -04:21:52.6  &    5.2    &    496.9 &                &     &                          91.5 \\ 
194     &    01:45:39.5 &  -04:34:36.8  &    7.4    &    496.4 &                &     &                          64.4 \\ 
195     &    01:45:00.9 &  -04:38:20.4  &    6.8    &    489.7 &                &     &                          69.8 \\ 
196     &    01:45:53.7 &  -04:27:07.9  &    6.3    &    483.2 &                &     &                          76.2 \\ 
197     &    01:44:58.9 &  -04:22:38.1  &    5.4    &    482.7 &                &     &                          85.8 \\ 
198     &    01:44:53.1 &  -04:46:41.4  &    5.3    &    481.7 &                &     &                          90.9 \\ 
199     &    01:45:37.9 &  -04:32:29.5  &    7.1    &    481.1 &                &     &                          65.2 \\ 
200     &    01:46:06.8 &  -04:39:41.9  &    6.0    &    477.5 &                &     &                          79.2 \\ 
201     &    01:45:00.7 &  -04:23:31.9  &    5.5    &    470.2 &                &     &                          84.8 \\ 
202     &    01:46:20.5 &  -04:28:20.7  &    5.1    &    468.7 &                &     &                          91.8 \\ 
203     &    01:46:03.1 &  -04:44:25.9  &    5.4    &    466.6 &                &     &                          84.5 \\ 
204     &    01:45:22.9 &  -04:45:24.8  &    5.9    &    464.3 &                &     &                          78.6 \\ 
205     &    01:45:12.2 &  -04:45:08.6  &    5.8    &    463.3 &                &     &                          78.3 \\ 
206     &    01:44:45.2 &  -04:42:18.4  &    5.3    &    451.0 &                &     &                          84.7 \\ 
207     &    01:45:52.0 &  -04:44:55.0  &    5.3    &    441.4 &                &     &                          82.5 \\ 
208     &    01:45:19.8 &  -04:44:15.2  &    5.7    &    438.9 &                &     &                          75.5 \\ 
209     &    01:44:59.3 &  -04:29:05.9  &    5.8    &    434.5 &                &     &                          73.0 \\ 
210     &    01:44:52.9 &  -04:31:50.3  &    5.7    &    431.1 &                &     &                          74.4 \\ 
211     &    01:46:07.8 &  -04:30:41.9  &    5.4    &    430.6 &                &     &                          79.8 \\ 
212     &    01:45:19.2 &  -04:41:31.3  &    5.9    &    430.3 &                &     &                          72.0 \\ 
213     &    01:45:46.9 &  -04:33:17.1  &    6.0    &    418.7 &                &     &                          69.8 \\ 
214     &    01:45:45.1 &  -04:26:05.9  &    5.3    &    409.6 &                &     &                          77.3 \\ 
215     &    01:45:58.4 &  -04:37:13.3  &    5.4    &    407.0 &                &     &                          74.4 \\ 
216     &    01:44:52.8 &  -04:34:19.1  &    5.3    &    403.4 &                &     &                          75.1 \\ 
217     &    01:45:46.7 &  -04:31:16.1  &    5.6    &    400.2 &                &     &                          68.1 \\ 
218     &    01:45:56.8 &  -04:29:02.2  &    5.0    &    389.5 &                &     &                          71.7 \\ 
219     &    01:45:15.9 &  -04:41:55.6  &    5.2    &    385.1 &                &     &                          74.2 \\ 
220     &    01:45:40.7 &  -04:43:16.6  &    5.0    &    385.0 &                &     &                          73.5 \\ 
221     &    01:45:48.1 &  -04:38:09.8  &    5.3    &    383.8 &                &     &                          71.0 \\ 
222     &    01:45:41.3 &  -04:39:41.1  &    5.2    &    374.9 &                &     &                          70.9 \\ 
223     &    01:44:58.0 &  -04:36:03.1  &    5.0    &    373.8 &                &     &                          73.8 \\ 
\end{tabular}
\end{table*}

\section{Source counts}
\label{section:sourcecounts}

The source counts at 1.4~GHz have been studied extensively (e.g. Windhorst \etal 1985;
Hopkins \etal 1998; Richards 2000; Seymour \etal 2004)
and show a flattening of the Euclidean normalised counts below $\sim$1 mJy.
This GMRT data reaches the depth required
to confirm or deny the existence of this `sub-mJy' bump at lower frequency.

\subsection{Completeness}
\label{section:complete}
To be 
compared meaningfully to other results and with models, several completeness corrections need to be made to 
convert the raw source numbers into a sky density as a function of flux density.

\subsubsection{Primary Beam Attenuation}
\label{section:pbat}
The dominant effect is the loss of sensitivity relative to that at the phase tracking centre due to the 
attenuation of the primary beam. This is corrected for in the binning process by weighting each source by the
area over which it could be detected, 
see Section \ref{section:construct}.

\subsubsection{Source Detection Completeness}
\label{section:noisebias}
Sources are detected based on peak flux density in a map of finite noise. The detected peak flux density of a
source is the superposition of its intrinsic peak flux density and any noise flux at its position. 
This effect can be sufficient to move sources in and out of the flux bins that correspond to their intrinsic fluxes. 
The important number in terms of completeness is the net gain or loss on the number of sources in each bin
and is known as Eddington bias. 
Since the source counts 
are rising steeply with decreasing flux, 
the number of low flux sources scattered up is likely to be greater that the number scattered down.  
To correct for this effect, 
some knowledge of the number of sources below the detection threshold is needed (Eddington 1913). Hence, we leave discussion of this 
Eddington bias, which is fairly small, until
section \ref{section:eddington}, after the modelling of the source counts is addressed and 
the source counts at at flux densities below the limit of our survey can be estimated.

\subsubsection{Instrumental Effects}
\label{section:intruments}
Bandwidth and time delay smearing reduce the peak flux density of a source
while conserving the total flux density. Such a reduction of peak flux
density can cause a source 
to be missed by a peak signal
source extraction. However, as was shown in Section \ref{section:corrections},
neither bandwidth nor time delay smearing is important
in the current study, and so no corrections have been made for these effects.

\subsubsection{Resolution Bias}
\label{section:resbias}
We are calculating source counts based on total flux density from a catalogue
selected on peak flux density, and so some sources of intrinsic low surface brightness
may be missed. 
As in previous deep surveys (Richards 2000,
Seymour \etal 2004) we have searched tapered, low-resolution, maps
for such sources and a small number have already been included in the 
catalogue. 

For any value of $S_{tot}$, an upper angular size limit exists whereby the peak flux density 
of a source of that total flux still exceeds the detection limit of the survey. 
Using the relation 
\begin{equation}
\frac{S_{tot}}{S_P} = \frac{\theta_{maj} \theta_{min}}{b_{maj} b_{min}}
\end{equation}
where $\theta_{maj}$ and $\theta_{min}$ are the major and minor axes of the 
detected source respectively and $b_{maj}$ and $b_{min}$ are the major and 
minor axes of the restoring beam (Prandoni \etal 2000b), we calculate the size limits relavent
to our full resolution and tapered maps.

\begin{figure}
\includegraphics[width=6cm,angle=0]{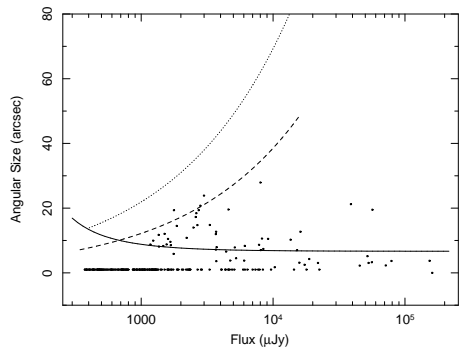}
\caption{
The angular size distribution of the sources in the 1$^H$ field. 
The solid line shows the minimum size for which a source can be confidently considered
resolved as a function of flux density (see Section \ref{section:resolution}).
The dashed and dotted lines show the maximum detectable size resolution bias limits
in the full resolution and tapered maps, respectively. All sources taken as unresolved
are arbitrarily plotted at 1 arcsec.
}
\label{fig:resbiasfig}
\end{figure}

At detected fluxes above 360$\mu$Jy, the tapered map provides excellent sensitivity to 
extended sources, where all sources smaller than 12 arcsec should be detected.
Further, very large sources will still be missed.
Due to the differing detection limits between the two maps, in the flux density 
range of 300 to 360$\mu$Jy, the smaller beam of the full resolution means that 
sources larger than $\sim$7 arcsec are missed.
Fig. \ref{fig:resbiasfig} shows the angular size distribution of the sources
in the catalogue, plotted against $S_{tot}$,
with the detectable size limits of both the full resolution and tapered maps 
plotted as dashed and dotted repectively.
Sources whose properties place them above these limits will not be detected in
our observations, and the number thereof consitutes the resolution bias.
As can be seen from the angular size distribution of the detected sources, 
few sources would be expected to reside in the zone where resolution bias
would exclude them from our catalogue. 

Little is known about the angular
size distribution of sub-mJy sources at 610~MHz, but the angular size distribution
of 
sources found in 
higher resolution deep surveys at 1.4~GHz over 
a similar flux range of 0.15 to 1mJy (the region in 610~MHz flux space for which 
resolution bias may have an effect, adjusted for 
a spectral index 
of
-0.7, and at sufficiently high flux that these 1.4~GHz surveys are not effected unduly
by resolution bias themselves) can provide some information about the expected 
population.
Bondi \etal (2003) fit the 1.4~GHz angular size distribution of the sources in the 
VLA-VIRMOS Deep Field over a flux range not expected to be affected by resolution bias
(0.4 to 1mJy). Using their formulation, we expect to miss $\sim$3 percent of the sources
above the tapered resolution bias limit in this flux range.
Assuming this angular size distribution holds to lower 1.4~GHz flux densities, we would expect to
miss only $\sim$7 percent of sources which lie between the detection limits of our full resolution and 
tapered 610~MHz maps.
Inspecting the
size distribution of sources in the 3.4~arcsec resolution VLA A+B array 1.4~GHz map of the
13$^H$ deep field (Seymour 2002) provides similar estimates of the fraction of sources that may be missed.
Hence, we conservatively estimate the effect of resolution bias as 7 
percent below a detected flux of 360$\mu$Jy, and 3 percent between 360$\mu$Jy and 1mJy.
It should be noted that the likely steep spectrum
of extended emission implies that the radius out to which an extended source could be
detected is greater at 610~MHz that at 1.4~GHz, implying larger angular sizes
at the lower frequency. 
Another limitation of examining the 1.4~GHz angular size
distribution is that higher resolution images can resolve out flux from
low surface brightness sources.
Nevertheless, the study of the 1.4~GHz size distribution as a guide to that at 610~MHz,
while not without problems, is informative.

\subsection{Construction of the Source Counts}
\label{section:construct}

We have binned our sources in total flux density with the lowest bin having 
a lower corrected flux limit of 344 $\mu$Jy.
Bins of width 0.3 dex, were chosen to provide good flux resolution 
while
retaining a reasonable number of sources per bin. The highest two flux bins 
are widened to 0.6 dex to compensate for the lower number of sources per decade
at high flux densities.

\begin{figure*}
\includegraphics[width=4in,angle=0]{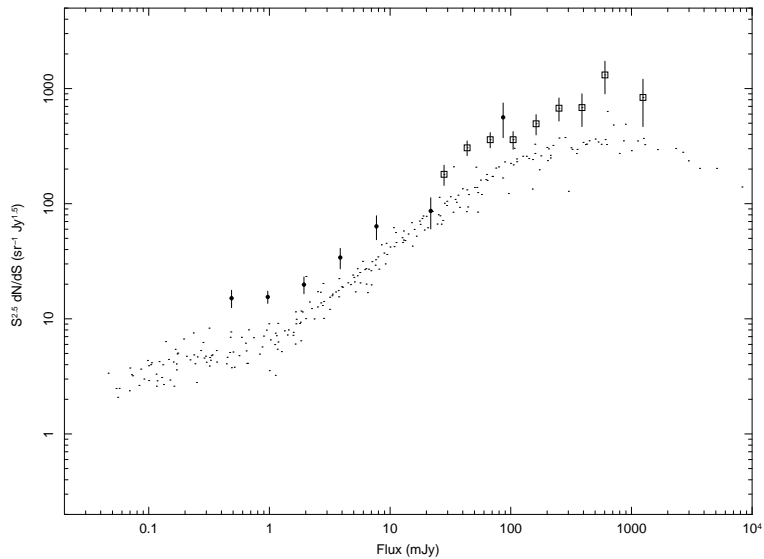}
\caption{The differential Euclidean-normalised 
610~MHz source counts as determined in the present study (filled circles), plotted at the log bin centre, $S_{mid}$,
and those of Katgert 1979 (empty squares) overlayed on those at 1.4
GHz (small dots) from the compilation of Seymour \etal 2004.}
\label{fig:610plus1p4}
\end{figure*}

Each source was weighted by the inverse of the area over which it could have been 
detected -- 
any departure from the full survey area being caused by primary beam attenuation and
the areas of high rms noise surrounding the bright sources 
-- and for the fraction of sources missed at that flux by resolution bias.
These weights were then summed for all the sources in each bin, 
To normalise to the counts expected from a Euclidean static universe,
the summed weight of each bin was divided by
$S_{mid}^{-2.5}$, where $S_{mid}$ is the 
flux density at the log centre of the bin.
The fractional uncertainty estimates are calculated from the inverse square root of the number of sources within 
the bin.
The bin flux limits, log bin centres, numbers of sources, 
and normalised differential source counts (which have been corrected for Eddington bias, see 
Section \ref{section:eddington})
are tabulated in Table \ref{tab:counts}, and plotted in
Fig \ref{fig:610plus1p4}.
Also plotted are the 610~MHz source counts determined from a number of Westerbork
survey fields of Katgert (1979), which can be seen
to agree well with the current observations. 
A prominent bump at sub-mJy levels, 
can indeed be seen at 610~MHz, where the
normalised counts flatten below $\sim2$~mJy.
Fig \ref{fig:610plus1p4} also includes a compilation of the source counts at 1.4~GHz (see Seymour
\etal 2004 and references therein). The counts at 610~MHz follow a similar shape to those at 1.4~GHz,
but with a higher normalisation 
consistent with the sources having, on average, a steep radio spectrum.
See Section \ref{section:model}
for a more detailed discussion of the relationship between the source counts in the two bands.

\begin{table}
\begin{tabular}{cccc}
Bin & $S_{mid}$ & $Number$ &    $S^{2.5}dN/dS$ \\
(mJy) & (mJy) &  &  (Jy$^{1.5}$sr$^{-1}$)  \\
\hline
0.344 -- 0.686  &  0.49  & 52 &   $15.1 \pm 2.6  $ \\  
0.686 -- 1.34  &  0.97  & 70 &    $15.6 \pm 1.9  $	\\  
1.34 -- 2.73  &  1.93  & 38 &    $20.2 \pm 3.2  $	\\  
2.73 -- 5.45  &  3.86  & 25 &    $34.4 \pm 6.9  $ \\  
5.45 -- 10.9  &  7.70  & 18 &    $63.6 \pm 15  $       \\  
10.9 -- 43.3  &  21.7  & 11 &    $86.5 \pm 26  $       \\  
43.3 -- 172  &  86.4  & 9 &    $562 \pm 187  $ \\  
\hline
\end{tabular}
\caption{Tabulated 610~MHz source counts. The columns show bin flux limits, 
log bin centre, number of sources, 
and Euclidean normalised dN/dS corrected for Eddington bias}
\label{tab:counts}
\end{table}

\subsection{Modelling the 610~MHz Source Counts}
\label{section:model}

The modelling of source counts has long been a method of constraining the cosmological evolution of radio 
populations, 
particularly those too optically faint to be spectroscopically identified 
(e.g. Longair 1966; Rowan-Robinson 1970; Hopkins \etal 1998;
Seymour \etal 2004, Huynh \etal 2005).
We model the 610~MHz source counts presented here following the 
method of Seymour \etal (2004), and refer the
reader to their section 4.3 for details of the process.
In brief, the local radio luminosity functions of both AGN and of star forming galaxies  
are modified by simple power law evolution in luminosity 
and integrated over redshift to predict the source counts expected
in deep radio surveys.

The AGN luminosity function and evolution function are taken from 
Rowan-Robinson \etal (1993) whose 1.4~GHz functions are based on survey 
work at 2.7~GHz by Dunlop \& Peacock (1990).
They split the local AGN population into two types, characterised by flat 
and steep radio spectra.
In this study, the steep spectrum luminosity function
was shifted to 610~MHz assuming an average spectral index of $\alpha = -0.7$. 
For the flat spectrum population ($\alpha = 0$) the lunimosity function at 610~MHz was 
taken as identical to that at 1.4~GHz.

The local starburst radio luminosity function was taken from Sadler \etal (2002), 
who take a sample
of 242 radio sources detected in the 
NVSS which are identified as starburst galaxies in the 2dF Galaxy Redshift Survey.
The Sadler \etal (2002) radio luminosity function was shifted to 610~MHz assuming an average spectral 
index of $\alpha = -0.7$, suitable
for the optically thin synchrotron radiation expected from supernova remnants associated with rapid
star formation.
Following previous authors, the luminosity evolution was parametrized as $(1+z)^Q$.
\begin{figure*}
\includegraphics[width=4in,angle=0]{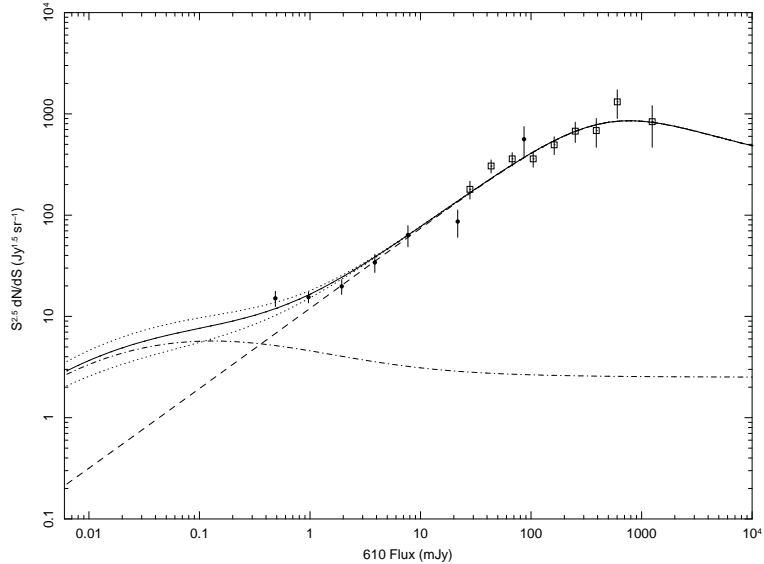}
\caption{Model fits superposed on the 610MHz source counts from Fig \ref{fig:610plus1p4}. 
The solid
line and accompanying dotted lines represent the best-fit total model 
and errors. 
The dashed and dot--dash lines represent the separate AGN and starburst galaxy contributions
respectively, as
discussed in the text. The AGN contribution is fixed but the luminosity of
the starburst contribution is allowed to evolve as $(1+z)^Q$.}
\label{fig:610models}
\end{figure*}

In order to fit this model to the 610~MHz counts, and obtain useful
errors on $Q$, we again follow previous
authors and fix the contribution to the counts of the AGN
populations. The 610~MHz source counts above $\sim$4 mJy are well
described by the model of the combined AGN population luminosity and
evolution functions as taken from Rowan-Robinson \etal (1993).  Also, we
assume zero density evolution, and hence the only parameter left free
to vary is the power law index of the starburst population luminosity
evolution ($Q$).  We have performed a $\chi^2$ fit, and the
best-fitting value is $Q=2.7^{+0.15}_{-0.25}$, where the errors are given at
the 1$\sigma$ level.  




\subsection{Eddington Bias and Detection Completeness}
\label{section:eddington}
When source counts are rising steeply with decreasing flux density 
it is likely that more sources will
be scattered 
by the noise towards brighter fluxes than to dimmer fluxes.
This effect is expected to be greatest at the flux limit of our 
survey.
To quantify the
effect that such Eddington bias has on source counts, 
we use the best-fitting population model of Section \ref{section:model} extrapolated to $60\mu$Jy
as a prediction 
of the 
source counts below the detection limit.
A population of fake sources with a
flux density distribution determined from 
this model 
were inserted
randomly into a residual noise map. 
The input flux density of each source was reduced, dependent upon position, to account for the attenuation of the 
primary beam sensitivity.
Sources were then extracted from these maps and
counts constructed from these detections, following the same method as used for 
the real
sources in Section \ref{section:construct}, and the process repeated 10 times.
The difference between the average 
detected differential source count and that of the input model
gives an estimate of the {\em net} up-scatter, i.e. the Eddington
bias.

Only the lowest flux bin shows a significant offset, with the average measured differential 
source count overestimating the inserted model value by 21 percent. 
In order to make a first-order approximation to the true source counts
we therefore reduced the observed counts in the lowest flux bin by
21 percent and refitted the model, obtaining a value of $Q=2.45$.
We note that the 21 percent correction factor is actually slightly larger
than we should be applying to return to the true source counts as it
is based on an overestimate of the real value of $Q$, corresponding to a steeper
unnormalised source count and hence to a larger number of up-scattered
sources than is the case in reality. 
However the 21 percent factor is not a
great deal too high as we can see by repeating the simulations
described above, assuming $Q=2.45$. In this case we find that the
simulated source counts exceeded the model, in the lowest flux bin, by
19 percent thereby almost, but not quite, returning us to our starting
position. 
Thus we adopt an
intermediate correction factor of 20 percent to the counts in the observed
lowest flux bin which, when refitted, still provides $Q=2.45^{+0.25}_{-0.4}$.
The corrected source counts and best fitting model are
shown in Fig. \ref{fig:610models}.

An f-test gives the probability of the improvement of fit 
upon the inclusion of the starburst component being statistically significant at the  
99.3 percent level.



\subsection{Comparison with other determinations}
\label{section:others}
At 1.4~GHz and following similar methodology, Rowan-Robinson \etal (1993) find $Q=2.5\pm0.5$,
Hopkins \etal (1998) find  $Q=3.1\pm0.8$,
Condon \etal (2002) find $Q=3.0\pm1$, Seymour \etal (2004) find $Q=2.5\pm0.5$, and Huynh \etal (2005)
find $Q=2.7$.
A study of IRAS sources believed to be tracing starforming galaxies by Saunders \etal (1990)
found evidence for luminosity evolution of $Q=3.2\pm0.1$.
Hopkins (2004) finds $Q=2.7\pm0.60$ and $P=0.15\pm0.60$ (where $P$ is the powerlaw index of 
density evolution) as 
the best-fitting evolutionary
scenario when both 1.4~GHz source counts and other star formation rate indicators are considered 
simultaneously.
The best-fitting value of $Q$ at 610~MHz ($2.45^{+0.25}_{-0.4}$) is therefore 
consistent, within errors, with 
the results from the plethora of measurements
at 1.4~GHz.
This agreement suggests that 
observations at both frequencies are probing similar source populations,
and that the sources which consitute
the 'sub-mJy bump'
can be explained by the an evolving population of starburst galaxies
with steep ($\alpha\sim-0.7$) radio spectra.

The basic two population model used here may over simplify the true
situation in faint radio sources. AGN and starburst activity can occur simultaneously
in the same galaxy, and their common requirement of gas reservoirs
would indicate that they are not entirely independent.
However, we have shown (Seymour \etal 2004, and this work) that such a model does provide a reasonably
good representation of the observed source counts at both 1.4~GHz and
610~MHz.

\section{Conclusions}
\label{section:conclusion}
In a $\sim$4.5 hour 610~MHz observation of the 1$^H$ \xmm/\ch 
$1^H$ survey area, 223 sources are detected down to a 5 $\sigma$ detection
limit of 300 $\mu$Jy within a 32 arcmin radius.

We confirm
that the 610~MHz source counts show the upturn in Euclidean
normalised differential counts below $\sim$2~mJy familiar from extensive studies at
1.4~GHz. The 610~MHz counts are reasonably well fit with a model similar to those fit
at 1.4~GHz, assuming a spectral index of $\alpha=-0.7$ for both
steep spectrum AGN and a population of starburst galaxies undergoing
luminosity evolution which dominate below $\sim1$ mJy.  
The best-fitting value for $Q$, the power law index of
the starburst luminosity evolution, is $2.45^{+0.25}_{-0.4}$. This is
consistent with the determinations at 1.4~GHz.

Future deeper and wider observations at 610~MHz,
will improve our understanding of the source counts at this frequency.
Future papers will consider the radio spectral properties of the sub-mJy
radio population which, combined with other multi-wavelength AGN/starburst discriminators
will allow the study of the separate evolution of the two populations.

We note that during the refereeing process, similar 610~MHz catalogues were presented 
by Bondi \etal (2006, astro-ph/0611704) and Garn \etal (2007, astro-ph/0701534), though
neither study models the populations seen at 610~MHz.

\section*{Acknowledgements}
DM acknowledges the support of a PPARC studentship, and IM$^c$H and TD acknowledge
support under PPARC grant PP/D001013/1.
We thank the staff of the GMRT for their very helpful support that made these observations possible.
GMRT is run by the National Centre for Radio Astrophysics of the Tata
Institute of Fundamental Research. DM and NS would like to thank Niruj
Mohan for indispensable advice and assistance.
The Digitized Sky Surveys were produced at the Space Telescope Science 
Institute under U.S. Government grant NAG W-2166.
We also thank Dave Green for the cutomised AIPS task UVFLP, and the referee, whose
comments greatly improved this manuscript.

\label{lastpage}

\bsp

\end{document}